\documentclass[12pt]{article}
\usepackage{graphicx}
\usepackage{amssymb}
\usepackage{amsmath}

\newcommand{\lsim}{ \mathop{}_{\textstyle \sim}^{\textstyle <} }
\newcommand{\bear}{\begin{array}}  
\newcommand {\eear}{\end{array}}
\newcommand{\bea}{\begin{eqnarray}}   
\newcommand{\eea}{\end{eqnarray}}
\newcommand{\beq}{\begin{equation}}   
\newcommand{\eeq}{\end{equation}}
\newcommand{\bef}{\begin{figure}}  \newcommand 
{\eef}{\end{figure}}
\newcommand{\bec}{\begin{center}}  \newcommand 
{\eec}{\end{center}}

\begin{document}

\begin{titlepage}

\begin{flushright}
ICRR-Report-566\\
\end{flushright}

\vskip 1.35cm

\begin{center}

{\large 
{\bf WIMP dark matter in gauge-mediated SUSY breaking  models \\
and its phenomenology} 
}

\vskip 1.2cm

Junji Hisano$^{a,b}$,
Kazunori Nakayama$^a$,
Shohei Sugiyama$^a$,
Tomohiro Takesako$^a$, \\
and
Masato Yamanaka$^a$

\vskip 0.4cm

{ \it $^a$Institute for Cosmic Ray Research,
University of Tokyo, Kashiwa 277-8582, Japan}\\
{\it $^b$Institute for the Physics and Mathematics of the Universe,
University of Tokyo, Kashiwa 277-8568, Japan}

\date{\today}

\begin{abstract} 
We propose an extended version of the gauge-mediated SUSY breaking models
where extra $SU(2)_L$ doublets and singlet field are introduced.
These fields are assumed to be parity-odd under an additional matter parity.
In this model, the lightest parity-odd particle among them would be dark 
matter in the Universe. In this paper, we discuss direct detection of the 
dark matter and the collider signatures of the model. 
\end{abstract}



\end{center}
\end{titlepage}

\section{Introduction}

Supersymmetry (SUSY) is one of the best-motivated physics beyond the standard 
model (SM) since it naturally solves the gauge-hierarchy problem.  SUSY must 
be spontaneously broken, however, in order for SUSY particles to obtain sizable 
masses. Among proposed SUSY breaking models, the gauge-mediated SUSY breaking 
(GMSB) models~\cite{Dine:1993yw} are interesting, since the SUSY flavor problem 
does not arise in GMSB models~\cite{Giudice:1998bp}.

Although low-energy phenomenology of GMSB models seems to be quite 
successful, it has non-trivial aspects from a cosmological point of view. In GMSB 
models the lightest SUSY particle (LSP) is the gravitino, superpartner of the 
graviton. The relic abundance of the gravitino in the Universe depends on the reheating 
temperature after inflation, and a stringent upper bound on the reheating 
temperature is obtained so that the gravitino does not exceed the present dark 
matter (DM) abundance~\cite{Moroi:1993mb}.
In particular, this implies that GMSB models are not compatible with the thermal 
leptogenesis scenario~\cite{Fukugita:1986hr} for most range of the gravitino 
mass. An exception is the low-energy GMSB models where the gravitino is 
lighter than about 10~eV and no upper bound on the reheating temperature is 
imposed~\cite{Viel:2005qj,Ichikawa:2009ir}. In this case, no candidate  for DM 
exists in the minimal setup of GMSB.\footnote{
In the low-energy GMSB scenario, a model in which a baryonic bound
state of strongly interacting messenger particles becomes cold DM is proposed
\cite{Dimopoulos:1996gy,Hamaguchi:2007rb}.
In this case DM has a mass of $ \mathcal{O}(100)$~TeV,
and cannot be detected by accelerator searches and direct detection experiments.
}

One may consider that the QCD axion~\cite{Kim:1986ax} can play a role of DM.
However, cosmology of SUSY axion models is quite non-trivial taking into 
account the existence of axino, fermionic superpartner of the axion, and saxion, 
scalar partner of the axion. The axino is produced thermally in the early Universe 
with a significant amount, and hence the reheating temperature is more severely 
restricted~\cite{Rajagopal:1990yx,Chun:1995hc,Covi:2001nw}.
The saxion coherent oscillation and its subsequent decay also gives catastrophic 
cosmological effects unless the reheating temperature is sufficiently low~\cite{
Asaka:1998xa,Kawasaki:2007mk}. Thus to make the axion the dominant 
component of DM in GMSB model requires careful considerations.\footnote{
  The strong CP problem may be solved in the low-energy GMSB models
  through the Nelson-Barr mechanism \cite{Nelson:1983zb,Barr:1984fh}.
  When the SUSY breaking sector is decoupled with the spontaneous
  CP-violating sector, the radiative correction to the QCD-theta term
  is suppressed due to the non-renormalization theorem.  Thus the
  axion is not necessarily needed in this case.  }

In this paper we extend the  GMSB models to include a 
candidate for WIMP DM. The minimum extension might be to add chiral 
supermultiplets with fundamental representation of $SU(2)_L$ ($H'$, $\bar H'$). 
Those fields are parity-odd under an additional $Z_2$-parity assigned. If all 
other minimal SUSY standard model (MSSM) fields are even under the $Z_2$-parity,
$H'$ and $\bar H'$ can be stable and a candidate for WIMP DM. 
Since they are weakly-interacting, their relic abundance falls into a correct range 
 in the thermal history of the Universe. However, this kind of extension is 
already excluded since it predicts too large direct detection rates in the DM search 
experiments through the coherent 
vector coupling to nucleons by $Z$-boson exchange. In order to avoid the direct 
detection bounds, we further introduce a singlet chiral multiplet $S'$ which is also
parity-odd. 
Then  WIMP DM becomes a mixture of $S'$, $H'$ and $\bar H'$, and the 
lightest parity-odd particle is real scalar boson or Majorana fermion so that the vector 
coupling of DM-nucleon is forbidden. Sizable interactions still exist through 
the Higgs exchange process, which is within the reach of on-going/future direct 
detection experiments.

This paper is organized as follows. In Sec.~\ref{sec:model} we define our model 
and study the properties of the DM particle. In Sec.~\ref{sec:direct} the DM direct 
detection rate is evaluated and it is shown 
that it is within the reach of current/future direct detection experiments. We discuss 
the LHC signature of this setup in Sec.~\ref{collider}. Finally in Sec.~\ref{sec:conc} 
we give conclusions.

\section{Model} \label{sec:model}

In this section we define our model and discuss the properties of the DM particles, 
such as their mass,  spin and interactions, in the model.
\begin{table}
  \caption{Particle contents of the model.}
  \label{table:par}
\begin{center}
\begin{tabular}{c|cccc}
 & $SU(3)_C$& $SU(2)_L$& $U(1)_Y$& $Z_2$\\ \hline\hline
$H$  & \bf{1}& \bf{2}& $-1/2$& even \\
$\bar H $ & \bf{1}& \bf{2}& $+1/2$& even \\
$H' $ & \bf{1}& \bf{2}& $-1/2$& odd \\
$\bar H' $ & \bf{1}& \bf{2}& $+1/2$& odd \\
$H'_c $ & $\bar{\bf{3}}$ & \bf{1} & $+1/3$ & odd \\
$\bar H'_c $ & $\bf{3}$ & \bf{1} & $-1/3$ & odd \\
$S'$& \bf{1}& \bf{1}& 0& odd \\
\end{tabular}
 \end{center}
\end{table}
We show parity-odd chiral multiplets newly introduced and MSSM Higgs doublets 
in Table~\ref{table:par}. The lightest particle among mixtures of parity-odd fields 
is stable and  can be a  WIMP DM candidate. In order to maintain the 
unification of the gauge couplings, we also introduce $SU(3)_C$ triplets ($H'_c$, 
$\bar H'_c$) which compose $\bf 5$ and $\bar{\bf 5}$ of $SU(5)$ with $H'$ 
and $\bar H'$. 
The most general renormalizable superpotential is
\begin{equation}\label{eq:superpotential} 
   \mathcal{L} 
   = -\int d^2 \theta \left(
   \mu H\bar H  + \mu' H'\bar H' + \mu'_c \bar H'_c H'_c
   +\lambda_1 H' \bar H S'+ \lambda_2 H\bar H' S' + \frac{1}{2}M_S S'^2
   \right)+\rm h.c. .
\end{equation}
$SU(2)_L$ products are defined as $H\bar H=H^0\bar H^0 -H^-\bar H^+$.  
Five real parameters and one CP violating phase, 
$\theta=\arg(\mu\mu' M_S\lambda_1^*\lambda_2^*)$, are introduced. 
For simplicity, we take $\theta=0$, so that
all parameters including $\mu'$ are real and positve.\footnote{
  The electric dipole moments (EDMs) are induced by electroweak
  two-loop diagrams in our model. In Ref.~\cite{Mahbubani:2005pt} the
  EDMs are discussed in a model similar to ours, where $SU(2)_L$
  doublet and singlet fermions are introduced in the standard
  model. It is found from their result that when the CP violating
  phase and the couplings are $\mathcal O(1)$, the electron EDM induced by
  parity-odd fermions reaches current experimental bound for
  $\mu'\sim M_S\lsim 1$~TeV. When $M_S\lsim \mu'$, the constraints are
  more loosened. Thus, the constraints are not severe at present,
  though future EDM searches might give severer constraints on the
  model or find the signature.
}

The $\mu$ parameter in Eq.~(\ref{eq:superpotential}) is of the electroweak scale 
from the naturalness argument. In addition, the mass parameters $\mu'$ and $M_S$ 
are also expected to be of the electroweak scale so that the DM relic abundance 
explains the WMAP measurement. Those mass parameters should have a common 
origin.  
In the low-energy GMSB models, an extra dynamical sector is introduced to
  generate a constant term in the superpotential for the cosmological constant 
to vanish. This sector can also produce the dimensional couplings in 
Eq.~(\ref{eq:superpotential}) with the magnitude of $\mathcal{O}(100)$~GeV 
\cite{Yanagida:1997yf}.\footnote{Fortunately,  the messengers do not generate
the $B$-terms in this mechanism at the leading order as mentioned in text, and we 
do not need to care about the so-called $\mu/B\mu$ problem. When the Higgs 
multiplets have (direct or indirect) couplings to SUSY-breaking field to generate the
$\mu$ term, the $B_\mu$ term would be too large.}

Soft SUSY-breaking terms for  the parity-odd fields are generically given as 
\begin{align}\label{eq:soft}
   V_{\rm soft} &= 
   m_{H'}^2|H'|^2+m_{\bar H'}^2|\bar H'|^2 + m_{S'}^2|S'|^2 \nonumber\\
   & + (B'_\mu H'\bar H' + \frac{1}{2}B'_S S'^2 + A^{\lambda}_1 H'\bar H S'
   +A^{\lambda}_2 H\bar H' S' +\rm h.c.).
\end{align}
We study particle spectrum based on the minimal gauge-mediation model (MGM)~\cite{mgm}
throughout this work, in which the messenger sector is composed of vector-like 
{\bf 5}+$\bar{\bf 5}$ representations of $SU(5)$. In the MGM, $A$- and $B$-terms 
vanish at the messenger scale $M_{\rm mess}$.  In this model, the scalar squared mass 
of the singlet is also zero at $M_{\rm mess}$. The only non-vanishing terms at $M_{\rm mess}$  
in Eq.~(\ref{eq:soft}) are
\begin{equation}\label{eq:m2h}
   m_{H'}^2=m_{\bar H'}^2=2N_{\rm mess} 
   \left[
   \frac{1}{4}\left(\frac{\alpha_Y}{4\pi}\right)^2
   + \frac{3}{4}\left(\frac{\alpha_2}{4\pi}\right)^2
   \right] \Lambda^2
   f \left(\frac{\Lambda}{M_{\rm mess}} \right),
\end{equation}
where  $N_{\rm mess}$ represents the number of $SU(5)$ representations introduced 
as messenger fields, $\Lambda=\langle F\rangle/M_{\rm mess}$, and $\langle F\rangle$ 
is the vacuum expectation value of the $F$-term which couples to messenger fields. 
The loop function $f(x)$ is given in Ref.~\cite{Giudice:1998bp}.

Now we discuss the properties of DM particles. The mass matrix for parity-odd fermions in this 
model is
\begin{equation}\label{eq:fermionic states}
  -\mathcal{L}= \frac{1}{2}\psi^T M_F\psi +\rm h.c. ,\quad
\psi=\begin{pmatrix}
      \tilde{S}' \\ \tilde{H}'^0 \\ \tilde{\bar{H}}'^0
     \end{pmatrix},
\end{equation}  
\begin{equation}
  M_F=\begin{pmatrix}\label{eq:MF}
       -M_S & -\lambda_1 \bar v & -\lambda_2 v \\
       -\lambda_1 \bar v & 0 & -\mu' \\
       -\lambda_2 v & -\mu' & 0
      \end{pmatrix},
\end{equation}
where $v=\langle H^0\rangle$, 
$\bar v=\langle \bar H^0\rangle$, 
and $v^2+\bar v^2=2m_Z^2/(g^2+g'^2)$.
Similarly, the squared mass matrices for parity-odd bosonic states are
\begin{equation}\label{bosonic states}
 -\mathcal{L}=\frac{1}{2}\varphi^T_R M_{B}^{(+)2} \varphi_R
+\frac{1}{2}\varphi^T_I M_{B}^{(-)2} \varphi_I ,\quad
 \varphi_R=
\begin{pmatrix}
 S'_R\\H'^0_R\\ \bar H'^0_R
\end{pmatrix}
,\,\varphi_I=
\begin{pmatrix}
 S'_I\\H'^0_I\\\bar H'^0_I
\end{pmatrix},
\end{equation}
\begin{equation}\label{eq:MB}
\scriptsize{
   M_B^{(\pm)2}=\begin{pmatrix}
 	      \lambda_1^2 \bar v^2 + \lambda_2^2 v^2 +M_S^2 +
		 m^2_{S'}\pm B'_S & \lambda_2\mu'v+\lambda_1 M_S 
		 \bar v\pm(\lambda_1\mu
		 v+A^{\lambda}_1\bar v) & \lambda_1\mu'\bar v
		 +\lambda_2 M_S v\pm(\lambda_2\mu\bar v+A^{\lambda}_2 v) \\
 	      \lambda_2\mu'v+\lambda_1 M_S \bar v\pm(\lambda_1\mu
		 v+A^{\lambda}_1\bar v)& \mu'^2 + \lambda_1^2\bar v^2
		 +m^2_{H'}+\Delta &\lambda_1\lambda_2 v \bar v\pm
		 B'_\mu  \\
 	      \lambda_1\mu'\bar v+\lambda_2 M_S v\pm(\lambda_2\mu
		\bar v+A^{\lambda}_2 v)& \lambda_1\lambda_2 v \bar v\pm
		 B'_\mu  & \mu'^2 + \lambda_2^2 v^2
		 +m^2_{\bar H'}+\bar \Delta
 	    \end{pmatrix}. 
}
\end{equation}
$\Delta$'s are contributions of $D$-term potential;
\begin{equation}\label{eq:D-term}
 \Delta=\frac{1}{2}m_Z^2\cos 2\beta,\quad \bar \Delta=-\Delta,
\end{equation} 
where $\tan\beta=\bar v/v$. The fields with subscripts $R$ and $I$ are CP-even 
and odd states, respectively, and canonically normalized as $\phi=\frac{1}{\sqrt{2}} 
(\phi_R +i\phi_I)$.

\begin{figure}[t]
 \begin{center}
   \includegraphics[width=0.7\linewidth]{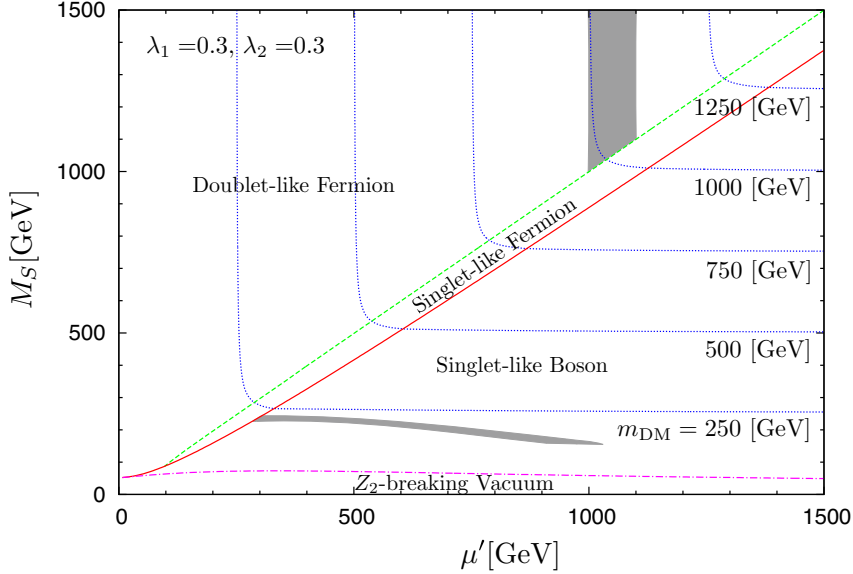} 
   \caption{Mass,  spin and type of the DM particle on $\mu'$-$M_S$~plane.
Dotted blue lines are contours of DM mass.  DM is boson
below the solid red line. It is fermion above the line.
DM is doublet-like on the left-hand side of the green broken line, 
and it is singlet-like on the right-hand side of the line.
The input parameters are given in text.
 Gray bands are regions where the relic DM abundance is consistent
  with the WMAP result within 2$\sigma$ level.
}
   \label{fig:mass}
 \end{center}
\end{figure}

We show mass,  spin and type of the lightest parity-odd particle in Fig.~\ref{fig:mass}. 
In the numerical calculation, we assume MGM with ${\bf 5}+\bar{\bf 5}$ messengers, 
and we use the result of \cite{Hisano:2007ah}. 
Here we took $\lambda_1 = \lambda_2 = 0.3$, $\tan\beta = 42$, $\mu = 660$ GeV, 
$N_{\rm mess}=1,$
the gluino mass 1 TeV, and $\Lambda/M_{\rm mess}=0.5$. $\mu$ and $\tan\beta$ 
are fixed from electroweak symmetry breaking conditions and a condition that $B_\mu=0$ 
at the messenger scale. We set these value as reference point throughout this work.

If $\mu'\gg M_S$, DM is singlet-like CP-even boson. The singlet bosons get no SUSY 
breaking mass terms in the GMSB model at the leading order. 
The bosonic and fermionic states are degenerate in masses. 
In our set up, the $F$-components of $H$ and $\bar{H}$ generate mass splitting between 
CP-even and odd states. Thus, one of the bosonic states tends to be
lighter than fermionic one. 
(If we take $\mu'$ negative and $|\mu'|\gg M_S$, DM is singlet-like 
CP-odd boson. In this case, mass splitting source of 
CP-odd boson is larger than that of CP-even boson as expected from
Eq.~(\ref{eq:MB}).)
If $M_S$ is too small, the lightest bosonic state becomes tachyonic and the $Z_2$-parity is broken 
spontaneously. If $\mu'\sim M_S$, DM is a fermion which is mixture of singlet and doublets. 
In the region of $\mu'\ll M_S$, DM is doublet-like fermion. The doublet-like bosonic states 
are heavier than fermionic one due to the GMSB effect.

  Now we see that this model predicts a correct relic DM abundance measured by 
WMAP~\cite{Komatsu:2010fb} with appropriate parameters.
Since our model includes particles with mass close to the DM mass, we should take into account 
the coannihilation effect for calculating the relic DM abundance.

First, consider the case where DM is singlet-like CP-even boson ($S'_R$).
In this case we consider three coannihilating states $(S'_R, S'_I,\tilde{S'})$.
The main annihilation processes in a non-relativistic limit are $S'_R S'_R \to h^0 h^0$ and 
$S'_I S'_I \to h^0h^0$, where $h^0$ stands for the SM-like Higgs boson, as far as they are 
kinematically allowed.
Here we can neglect other processes like $\tilde{S'} \tilde{S'} \to h^0 h^0$,
$S'_R \tilde{S'} \to h^0 \tilde{H^0}$, and $S'_R S'_I \to h^0 A^0$ where $A^0$ is CP-odd Higgs boson, and so on, because $\tilde H^0$ and $A^0$ are assumed to be heavy.
The reason for omitting the $\tilde{S'}$ annihilation is as follows. Since $\tilde{S'}$ 
is a Majorana fermion, in the non-relativistic limit, a pair of them is in CP-odd state. Then the opening channel from the viewpoint of CP 
conservation is the pair annihilation of $\tilde{S'}$ into a pair of $h^0$ in $p$-wave 
state. This channel, however, is forbidden by the angular momentum conservation.

The DM effective cross section~\cite{Griest:1991ma} of these processes is given by 
\begin{equation}
	\langle \sigma _{\mathrm{eff}} v \rangle
	= \left( \frac{1}{1+w} \right) ^2 \langle \sigma v\rangle_{S'_RS'_R \to h^0h^0}
	+ \left( \frac{w}{1+w} \right) ^2 \langle \sigma v\rangle_{S'_IS'_I \to h^0h^0} ,
\end{equation}
where
\begin{gather}
\langle \sigma v\rangle_{S'_RS'_R \to h^0h^0}
	= \frac{1}{64\pi} \frac{1}{\bar m_{S'_R}^2}\sqrt{1-\left ( \frac{m_{h^0}}{\bar m_{S'_R}} \right )^2}
	\left (\lambda_1^2
	- \frac{(M_S\lambda_1+A_1^\lambda)^2}{\bar m_{H_1'}^2+\bar m_{S'_R}^2-m_{h^0}^2}
	- \frac{(\mu'\lambda_1+\mu\lambda_2)^2}{\bar m_{H_2'}^2+\bar m_{S'_R}^2-m_{h^0}^2}
	\right )^2,\\
\langle \sigma v\rangle_{S'_IS'_I \to h^0h^0}
	= \frac{1}{64\pi} \frac{1}{\bar m_{S'_I}^2}\sqrt{1-\left ( \frac{m_{h^0}}{\bar m_{S'_I}} \right )^2}
	\left (\lambda_1^2
	- \frac{(M_S\lambda_1-A_1^\lambda)^2}{\bar m_{H_1'}^2+\bar m_{S'_I}^2-m_{h^0}^2}
	- \frac{(\mu'\lambda_1-\mu\lambda_2)^2}{\bar m_{H_2'}^2+\bar m_{S'_I}^2-m_{h^0}^2}
	\right )^2,\\
w = \left(\frac{\bar m_{S'_I}}{\bar m_{S'_R}} \right)^{\frac{3}{2}}\exp 
\left( -x \frac{\bar m_{S'_I} - \bar m_{S'_R}}{\bar m_{S'_R}} \right),\\
x = \frac{\bar m_{S'_R}}{T_{\mathrm{freeze~out}}} .
\end{gather}
Here, $\bar m_{S'_R}$, $\bar m_{S'_I}$, $\bar m_{H'_1}$, $\bar m_{H'_2}$ are the mass eigenvalues
and $T_{\mathrm{freeze~out}}$ is the freeze-out temperature of $S'_R$.
The relic DM abundance, in terms of the density parameter  $\Omega_{S'_R} h^2$, is expressed
as
\begin{equation}
	\Omega_{S'_R} h^2  
	=  0.10\left( \frac{2.7 \times 10^{-9} ~\mathrm{GeV^{-2}}}{\langle \sigma _{\mathrm{eff}} v \rangle} \right).
\end{equation}

The gray region in Fig.~1 (bottom one) shows the parameter space in which the DM relic abundance is 
consistent with WMAP observations within 2$\sigma$ level \cite{Komatsu:2010fb}. 
Thus we can see that the relic 
abundance falls into the correct range measured by WMAP for certain choice of parameters. 
We note that the coannihilation effect can change the relic abundance only 10$\%$ or so in 
the correct range for DM abundance.@

In the case of doublet-like fermion DM, we consider four coannihilating states ( $\tilde H'^{0}, 
\tilde {\bar H}'^{0}, \tilde H'^{-}, \tilde {\bar H}'^{+} $ ). The cross section and its relic 
abundance are basically the same as those of MSSM Higgsino-like DM whose mass is heavier than 
$W$-boson mass. If the coannihilation effect is efficient, the DM effective  cross section and its 
relic abundance are given in Ref.~\cite{Arkani-Hamed:2006fe};
\begin{gather}
	\langle \sigma _{\mathrm{eff}} v \rangle \label{eq:Arkani}
	= \frac{g^4}{512 \pi \mu'^{2}} \left( 21 + 3 \tan^2\theta _W +
 11 \tan^4 \theta _W \right), \\
\Omega_{\tilde H'^{0}} h^2  \label{eq:abun}
	=  0.10 \left( \frac{\mu '}{1~\mathrm{TeV}} \right) ^2,
\end{gather}
where four states are taken to be  degenerate in mass.\footnote{
The $t \bar t$ final state via stop exchange is ignored in 
Ref.~\cite{Arkani-Hamed:2006fe}, and our 
case does not include such a process. 
}
The effective cross section in Eq.~(\ref {eq:Arkani}) drops final states including 
ordinary MSSM superpartners, because their total contribution is at most 10$\%$ 
in our reference point  comparing with the SM final state contribution. The correct 
DM relic abundance consistent with WMAP observations within 2$\sigma$ level 
is  obtained in the gray region in Fig.~1 (top one).

Finally, we discuss about $H'_c$ and $\bar H'_c$ in
Eq.~(\ref{eq:superpotential}),  which are introduced in order to
maintain the successful gauge coupling unification. They have color
and electric charges, and their masses are also expected to be
order of electroweak scale.\footnote{
 If $\mu'_c$ is equal to
$\mu'$ at the GUT scale, $\mu'_c$ is about twice larger than
$\mu'$ at the electroweak scale.
} 
In the model, they are stable at the renormalizable level and
unstable due to the higher dimensional operators suppressed by 
relevant physics scale $M_{\text{phys}}$ such as 
\begin{equation}
 \begin{split}
   \frac{1}{M_{\text{phys}}} ({\bf 5 \bar 5_{SM}}) ({\bf 5 \bar 5_{SM}}) .
 \end{split}     \label{5 dim.ope.}
\end{equation}
In the GMSB model, triplet fermion is lighter than bosonic one. When
DM is doublet-like fermion, the lifetime is $\mathcal O(1) -  \mathcal O(10^3)$~sec for
masses $100-1000$ GeV in the case that ${M_{\text{phys}}}= M_{\rm GUT}
\simeq 10^{16}$ GeV. Here, it is assumed that the triplet fermion can
decay into the MSSM SUSY particles directly. When DM is singlet-like
boson, the decay of the triplet fermion is suppressed by the mixing
between the singlet and doublet states.  The feature of these
long-lived colored particles are constrained from the viewpoint of
cosmology. Their relic abundance would be determined by geometrical
cross sections of order of 10~mb effectively at temperatures below the
QCD phase transition, and they annihilate efficiently before the Big
Bang nucleosynthesis (BBN) if their masses are lighter than about
$10^{11}$~GeV~\cite{Kang:2006yd}. Their relic abundance after this
annihilation is estimated as $10^{-8}n_b(\mu'_c/{\rm TeV})^{1/2}$,
where $n_b$ is the baryon number density.\footnote{
It is also argued in
Ref.~\cite{Kusakabe:2009jt} that even such a small abundance may
affect the BBN and a lifetime of the colored particles should be
shorter than about 200~sec. This would give a lower bound on the 
triplet fermion mass, depending on 
$M_{\text{phys}}$ and the main decay mode, though we do not include this constraint in this paper.
}
The future collider experiments may prove existence of the colored particles.  
We discuss its possibility in Sec.~\ref{collider}.

\section{Direct detection rates} \label{sec:direct}

In this section we evaluate the DM-nucleon scattering cross section in our model. 
DM particles in the model have sizable interactions with nucleons and hence
they may be detected through on-going or future direct detection experiments.

First let us consider the case where the DM particle is fermionic (denoted by $\tilde \chi$).
DM interacts with nucleons through the $Z$-boson and Higgs-boson exchange diagrams.
The former yields a spin-dependent (SD) and 
the latter yields a spin-independent (SI) effective interactions.
The effective Lagrangian for these interactions is written at the parton level as
\begin{equation}
 \mathcal{L}_{\rm eff}=\sum_{q=u,d,s}d_q
 \bar{\tilde{\chi}}\gamma^\mu\gamma_5\tilde{\chi}
  \bar q \gamma_\mu\gamma_5 q 
 +\sum_{q=u,d,s,c,b,t}
f_q\bar{\tilde{\chi}}\tilde{\chi}\bar q q ,
\end{equation}
where coupling constants are given by
\begin{equation}
 d_q=\frac{g^2 T_{3q}}{8m_W^2}\left((O_F)_{12}^2-(O_F)_{13}^2\right),
\end{equation}
and
\begin{equation}
 f_q=m_q\frac{g}{2m_W}\left(
\frac{c_{h\tilde{\chi}\tilde{\chi}}c_{hqq}}{m_{h^0}^2}
+\frac{c_{H\tilde{\chi}\tilde{\chi}}c_{Hqq}}{m_{H^0}^2}
\right),   \label{fq}
\end{equation}
with
\begin{equation}
 c_{huu}=\frac{\cos\alpha}{\sin\beta},\quad 
 c_{Huu}=\frac{\sin\alpha}{\sin\beta},
\end{equation}
\begin{equation}
 c_{hdd}=-\frac{\sin\alpha}{\cos\beta},\quad 
 c_{Hdd}=\frac{\cos\alpha}{\cos\beta},
\end{equation}
\begin{equation}
  c_{h\tilde{\chi}\tilde{\chi}}=\frac{1}{\sqrt{2}}
\left(\lambda_1(O_F)_{11}(O_F)_{12}\cos\alpha
-\lambda_2(O_F)_{11}(O_F)_{13}\sin\alpha\right),
\end{equation}
\begin{equation}
  c_{H\tilde{\chi}\tilde{\chi}}=\frac{1}{\sqrt{2}}
\left(\lambda_1(O_F)_{11}(O_F)_{12}\sin\alpha
+\lambda_2(O_F)_{11}(O_F)_{13}\cos\alpha\right).
\end{equation}
Here $O_F$ is a $3\times 3$ mass diagonalizing matrix, which is obtained from 
Eq.~(\ref{eq:MF}).

Using these couplings, the SI scattering cross section between DM and nucleus with 
mass $m_T$ is expressed as \cite{Drees:1993bu,Jungman:1995df}
\begin{equation}
 \sigma_{\rm SI}=\frac{4}{\pi}\left(\frac{m_{\tilde{\chi}} m_T}
{m_{\tilde{\chi}}+m_T}\right)^2(n_p f_p+n_n f_n)^2,
\end{equation}
and the SD cross section is given by
\begin{equation}
 \sigma_{\rm SD}=\frac{4}{\pi}\left(\frac{m_{\tilde{\chi}} m_T}
{m_{\tilde{\chi}}+m_T}\right)^2\left[
4\frac{J+1}{J}(a_p\langle S_p\rangle + a_n\langle S_n\rangle)^2\right].
\end{equation}
Here $n_p (n_n)$ is the number of proton (neutron) in the target nucleus.  $J$ is the 
total nuclear spin, and $\langle S_{p(n)} \rangle = \langle A | S_{p(n)} |A \rangle$ 
are the expectation values of the spin content of the proton and neutron groups within 
the nucleus $A$~\cite{Engel:1989ix}, and  
\begin{equation}
 a_N=\sum_{q=u,d,s}d_q\Delta q_N ,
\end{equation}
\begin{equation}
 2s_\mu\Delta q_N=\langle N|\bar q \gamma_\mu\gamma_5 q|N\rangle.
\end{equation}
where $s_\mu$ is the nucleon's spin and $N=n,p$.
The DM-nucleon effective coupling is constructed from 
the DM-quark effective coupling as follows~\cite{Shifman:1978zn},
\begin{equation}
 \frac{f_N}{m_N}=\sum_{q=u,d,s}\frac{f_q f_{T_q}^{(N)}}{m_q}
 +\frac{2}{27}f_{T_G}\sum_{q=c,b,t}\frac{f_q}{m_q},    \label{f_N}
\end{equation}
\begin{equation}
 f_{T_G}=1-\sum_{q=u,d,s}f_{T_q}^{(N)}.
\end{equation}
For the nucleon mass matrix elements, we take $f_{T_u}^{(p)} = 0.023$, $f_{T_d}^{(p)} = 0.034$, 
$f_{T_u}^{(n)} = 0.019$, $f_{T_d}^{(n)} = 0.041$~\cite{Gasser:1990ce, Adams:1995ufa}
and $f_{T_s}^{(p)} = f_{T_s}^{(n)} = 0.025$~\cite{Ohki:2008ff}. 

Next let us consider the case of bosonic DM ($\chi$).
The effective Lagrangian through the Higgs boson exchange diagram is written as
\begin{equation}
 \mathcal{L}_{\rm eff}=\sum_{q=u,d,s,c,b,t}
 f_q\chi^2\bar q q.
\end{equation}
This yields the following SI scattering cross section,
\begin{equation}
 \sigma_{\rm SI}=\frac{1}{\pi m_{\chi}^2}\left(\frac{m_{\chi} m_T}
{m_{\chi}+m_T}\right)^2 (n_p f_p+n_n f_n)^2,
\end{equation}
where $f_p (f_n)$ is given by Eq.~(\ref{f_N}), with 
$c_{h\tilde \chi \tilde \chi} (c_{H\tilde \chi \tilde \chi})$ in $f_q$ (Eq.~(\ref{fq}))
is replaced by  the following,
\begin{equation}
 c_{h\chi\chi}=\frac{1}{2\sqrt{2}}\left(
(O_B Y_1 O_B^T)_{11}\sin\alpha - (O_B Y_2 O_B^T)_{11}\cos\alpha
\right),
\end{equation}
\begin{equation}
 c_{H\chi\chi}=\frac{1}{2\sqrt{2}}\left(
-(O_B Y_1 O_B^T)_{11}\cos\alpha - (O_B Y_2 O_B^T)_{11}\sin\alpha
\right),
\end{equation}
\begin{equation}
 Y_1=\frac{\partial}{\partial v}M_B^2,\quad
 Y_2=\frac{\partial}{\partial \bar v}M_B^2.
\end{equation}
Here $O_B$ is a $3\times 3$ mass diagonalizing matrix for the mass matrix 
$M_B^2(\equiv M_B^{(+)2})$, which is given in Eq.~(\ref{eq:MB}).



\begin{figure}
 \begin{center}
\begin{tabular}{c}
   \includegraphics[width=0.7\linewidth]{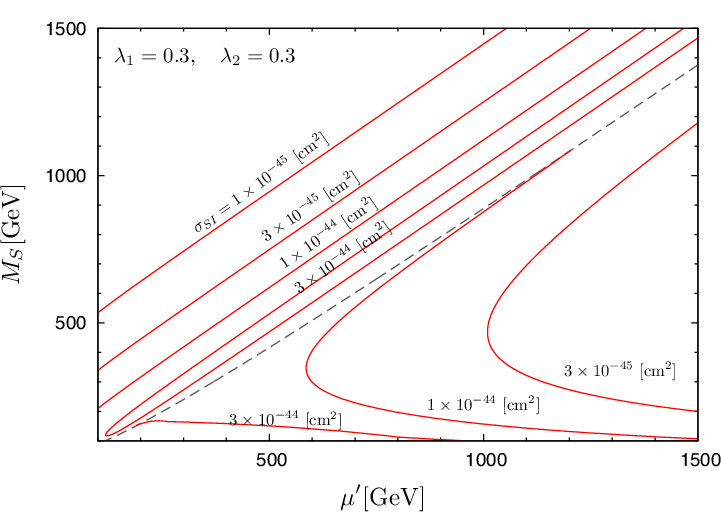}\\
   \includegraphics[width=0.7\linewidth]{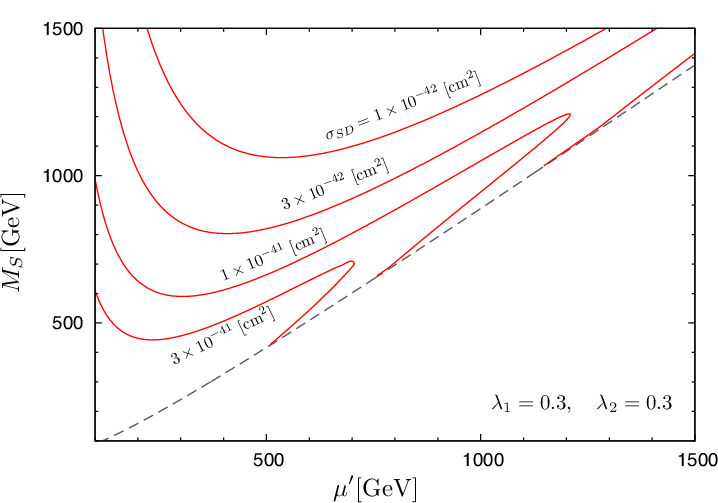}
\end{tabular}
   \caption{Contours of SI (top) and SD (bottom) cross sections with 
   a proton on $\mu'$-$M_S$ plane. Broken lines are boundaries 
   between fermionic and bosonic DM. }
   \label{fig:cross section}
 \end{center}
\end{figure}

Fig.~\ref{fig:cross section} shows contours of the SI and SD interactions with a 
proton on $\mu'$-$M_{S}$ plane. Parameters are set to be the same as those in 
Fig.~\ref{fig:mass}. The most stringent bound on the SI cross section comes 
from CDMS-II results~\cite{Ahmed:2008eu,Ahmed:2009zw}. The bound reads 
$\sigma_{\rm SI}/m_\chi \lesssim 3\times 10^{-46}~{\rm cm^2/GeV}$ for 
$m_\chi \gtrsim 100~$GeV. The observed two DM-like events at 
CDMS-II~\cite{Ahmed:2009zw} can be explained for appropriate parameters, 
if we take the two events seriously and assume they are caused by DM-nucleon 
scatterings. The predicted SI cross section is within the reach of future or on-going 
direct detection experiments.

Among the direct detection experiments, the best bound on the SD cross section 
comes from the XENON experiment \cite{Angle:2008we} ($\sigma_{\rm SD}/
m_\chi \lesssim 3\times 10^{-40}~{\rm cm^2/GeV}$ for $m_\chi \gtrsim 
100~$GeV).  In addition, the SD cross section is also constrained by observations 
of energetic neutrinos from the Sun produced by annihilations of DM particles 
captured by the Sun \cite{Griest:1986yu,Ritz:1987mh,Kamionkowski:1991nj,
Jungman:1994jr}.
Super-Kamiokande~\cite{Desai:2004pq}, AMANDA~\cite{Ackermann:2005fr} 
and IceCube with 22 strings give the stringent limits~\cite{Abbasi:2009uz}. 
The fermionic DM in the present model mainly annihilates into $W^+W^-$, and 
their subsequent decay modes $W \to e\nu$ produce high-energy neutrinos 
which may be detectable at IceCube. The current IceCube bound for mass at 
250 GeV is $\sigma_{\rm SD}\lsim 3\times 10^{-40}~{\rm cm^2}$, assuming 
the $W^+W^-$ final state.

 The bound on the SD cross section is still far from the
  prediction. However, the experiments sensitive to it would be
  important to determine spin of the DM particle after the DM is
  discovered. Furthermore, we might reconstruct the model by using
  results of the direct DM searches sensitive to SD and SI cross
  sections under assumption of  the
  thermal relic scenario,
 when the DM is fermionic.  This is because the
  model has only four input parameters, $\mu'$, $M_S$ and
  $\lambda_{1/2}$.

\section{Collider signatures}   \label{collider} 

One of the handles to confirm this model is discovery for the signatures of 
the extra particles at the LHC. In the confirmation, the key issue is the 
selection of missing energy events of this model under the backgrounds of 
ordinary GMSB models. In this section, we estimate the discovery reach of 
extra  parity-odd particles, and discuss the feasibilities for the selection 
assuming that the GMSB model has been already experimentally established 
as the SUSY breaking model.

\begin{figure}[t!]
  \begin{center}
   \includegraphics[width=305pt, clip]{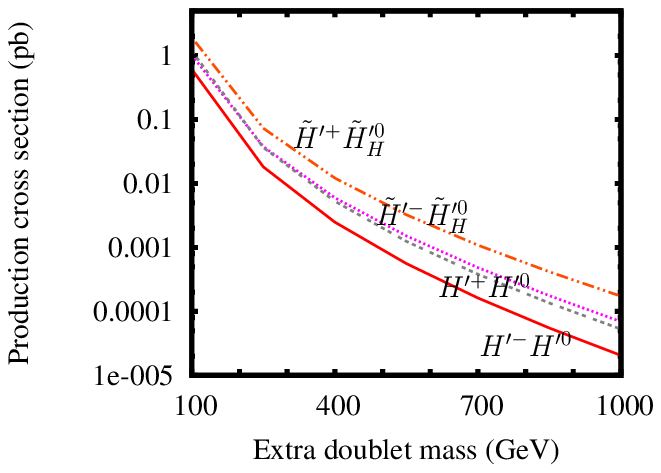}
   \end{center}
   \caption{{\small{Cross sections for $pp \rightarrow H'^0 H'^{\pm}$ 
   and $pp \rightarrow \tilde H'^0_H \tilde H'^{\pm}$. Here $H'^0$ and   
   $H'^{\pm}$ stand for neutral CP-even (CP-odd) and charged extra 
   boson, which are components of common extra doublet. 
   $\tilde H'^{0}_H$ and $\tilde H'^{\pm}$ stand for superpartners of 
   neutral and charged extra doublet boson.  
   We took only direct pair production as their production processes, and 
   included all combinations of final state pair in each result.}} }
   \label{production}
   \begin{center}
   \includegraphics[width=305pt, clip]{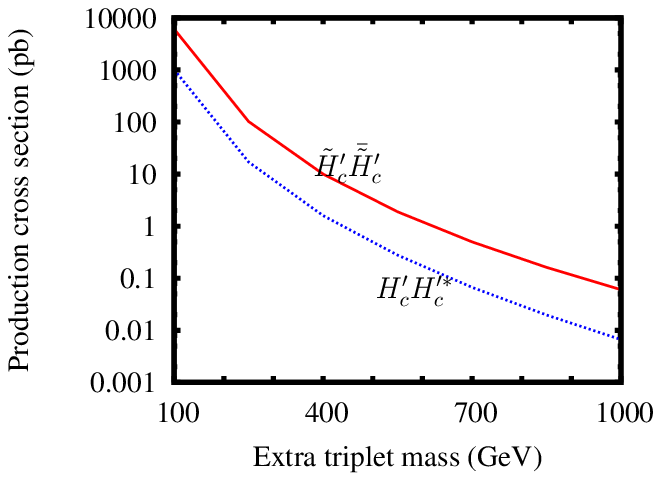}
   \end{center}
   \caption{{\small{ Production cross sections for extra triplet boson 
   $H'_c$ and its superpartner $\tilde H'_c$ as a function of their mass. 
   We took only direct pair production as their production processes.}} }
   \label{colored}
\end{figure}

First, we discuss the case that the lightest parity-odd particle is doublet-like 
neutral fermion, $\tilde H_L'^0$. The mass difference between 
$\tilde H_L'^0$ and heavier doublet-like neutral (charged) fermion, 
$\tilde H_H'^0$ ($\tilde H'^\pm$), is less than 40 GeV when 
$\mu'>94$~GeV within the parameters of Fig.~\ref{fig:mass}. Accordingly,  
the two-body decays of $\tilde H_H'^0$ and $\tilde H'^\pm$ are inaccessible.

Clean multilepton events, $pp \rightarrow \tilde H_H'^0 \tilde H'^\pm 
\rightarrow l \bar l \tilde H_L'^0 + l' \nu_{l'} \tilde H_L'^0$, offer a 
promising way for the selection. In the low-energy GMSB models the
SUSY events accompany energetic tau leptons or photons
since the next-lightest SUSY particle is typically stau or Bino-like  
neutralino. 
Thus, an observation of the missing energy events accompanying trileptons 
clearly point to $\tilde H_H' \tilde H'^\pm$ production as its source and 
would be the evidence of underlying physics responsible for the extra particles.

In order to optimize the trileptons events, one should reduce the $W^\pm Z$ 
background and $t \bar t$ background with suitable cuts. Those have 
been discussed in works \cite{Baer:1994nr, Baer:1995va, Aad:2009wy}, 
which focus on trileptons events from chargino-neutralino production, and 
after the cuts they find a total SM background of 19.6~fb. 
Here cuts designed in Ref~\cite{Aad:2009wy} are as follows: 
\begin{itemize}
   \item[1.] 3 leptons with $p_T > 10$ GeV. 
   \item[2.] At least one Opposite Sign Same Flavor (OSSF) dilepton with 
                   $M_{\text{OSSF}} > 20$ GeV to suppress low-mass 
                   $\gamma^*,~J/\Psi,~\Upsilon$, and conversion backgrounds. 
   \item[3.] Lepton track isolation: $p_{T, \mathrm{trk}}^{0.2} < 1$ GeV for muon 
                   and $< 2$ GeV for electron, where $p_{T, \mathrm{trk}}^{0.2}$ is the 
                   maximum $p_T$ of any additional track within a cone $R$ = 0.2 
                   around the lepton. 
   \item[4.] $E_T^{\mathrm{miss}} > 30$ GeV. 
   \item[5.] No OSSF dilepton has invariant mass in the range 81.2 GeV $< 
                   M_{\text{OSSF}} <$ 102.2 GeV. 
   \item[6.] No jet with $p_T > 20$ GeV.                
\end{itemize}
In addition to the SM background, we also comment on the background coming 
from ordinary superpartners. The GMSB model with stau NLSP may produce the 
trilepton events, e.g., $\tilde \chi_1^\pm \tilde \chi_2^0 \to \tilde 
\tau \bar \nu_\tau + \tilde \tau \bar \tau \to \tilde G \tau \bar \nu_\tau 
+ \tilde G \tau \bar \tau \to E_T^{\mathrm{miss}} +$ trilepton, and so on.  Those MSSM 
backgrounds include tiny branching ratio BR($3 \tau \to l + \text{OSSF dilepton} 
+ E_T^{\mathrm{miss}}$), and hence the MSSM backgrounds could be reduced enough.

Red dash-dotted and purple dotted lines in Fig.~\ref{production} show the 
cross sections for the direct production $pp \rightarrow \tilde H_H'^0 \tilde 
H'^+$ and $pp \rightarrow \tilde H_H'^0 \tilde H'^-$, respectively. They are 
calculated with the CalcHEP \cite{Pukhov:2004ca} implementing the Lagrangian 
Eq.~(\ref{eq:superpotential}) and the CTEQ6L code \cite{Pumplin:2002vw} 
as a parton distribution function. In the calculation, we took $m_{\tilde 
H_H'^0} = m_{\tilde H'^\pm}$, and set the center of mass energy to be 
$\sqrt{s} = 14$ TeV. 
From the numerical result, production cross section can be parametrized as 
$\sigma (pp \rightarrow \tilde H_H'^0 \tilde H'^\pm) = 2.47 \times 
10^{-4}(\text{TeV}/m_{\tilde H})^{4.1}$ pb,  where $ 
m_{\tilde H_H'^0} = m_{\tilde H'^\pm}(\equiv m_{\tilde H'})$.

Since their decay modes into 
two bodies are kinematically inaccessible, they decay into dileptons and  $\tilde 
H_L'^0$ via off-shell weak gauge boson. Their branching ratios into dileptons 
are, therefore, uniquely determined; BR$(\tilde H_H'^0 \rightarrow l \bar l 
\tilde H_L'^0) \simeq  6.73 \%$ and BR$(\tilde H'^\pm \rightarrow l' 
\nu \tilde H_L'^0) \simeq 21.32 \%$. Here results have summed over 
electron and muon. 
Thus the $5\sigma$ discovery reach for the trilepton signals is estimated as follows, 
\begin{equation}
 \begin{split}
   N_{\text{signal}} &= 
   \sigma (pp \rightarrow \tilde H_H'^0 \tilde H'^\pm) \times \mathcal{L} 
   \times \text{BR} (\tilde H_H'^0 \rightarrow l \bar l \tilde H_L'^0) 
   \times \text{BR} (\tilde H'^\pm \rightarrow l' \nu \tilde H_L'^0)   \\
   &\simeq 0.354 \times \left( \frac{\text{TeV}}{m_{\tilde H'}} \right)^{4.1} 
   \left( \frac{\mathcal{L}}{100 \text{ fb}^{-1}} \right)
   > 5 \sqrt{N_{\text{BG}}} ~.
 \end{split}     \label{5sigma_1}
\end{equation}
Here $\mathcal{L}$ stands for an integrated luminosity, and 
$N_{\text{signal}}$ ($N_{\text{BG}}$) stands for the number of multilepton 
events (SM background events).  Thus, assuming an integrated luminosity 100 
fb$^{-1}$ and demanding the SM background 19.6~fb , 
the superpartners of extra doublet for $m_{\tilde H_H'^0} = m_{\tilde H'^\pm} 
\lesssim$ 205~GeV would be discovered at the 5$\sigma$ level. 
Indeed, for the discussion of discovery reach, we should mention the 
acceptance of detector. However it needs precise simulation of signal events, 
and is beyond the scope of this work. 


Next, we discuss the case that the lightest extra particle is singlet-like 
boson, $S'$. They are mainly produced by the decay of heavier 
doublet-like bosons accompanying the SM-like Higgs boson, $h^0$.  
Accordingly, the signal events contain $b \bar b$ and the missing energy, 
and hence provide a distinguishable signature from the ordinary GMSB ones.

The dominant background to the two $b$-jets plus large missing energy events 
presumably comes from $t \bar t$ production. It can be reduced by the 
following cuts:
\begin{itemize}
   \item[1.] $E_T^{\mathrm{miss}} >$ 100 GeV.
   \item[2.] $b$-jets with $p_T > $ 50 GeV.
   \item[3.] $E_T^{\mathrm{miss}} + \sum E_{Tj} >$ 1500 GeV.            
\end{itemize}
Here $\sum E_{Tj}$ indicates the transverse energy sum over untagged 
jets~\cite{Baer:2000pe}. The most promising event for the discrimination, 
therefore, would be missing energy events accompanying $b \bar b$ and 
energetic jet, $pp \rightarrow H_H'^0 H'^\pm \rightarrow h^0 S' + W^\pm 
S' \rightarrow b \bar b S' + \text{jet}S'$. 
In both GMSB models with stau NLSP and Bino-like neutralino NLSP, there 
would exists no background events of ordinary superpartners under the cut conditions.

Gray dashed and red solid lines in Fig.~\ref{production} show the cross 
sections for the direct production $pp \rightarrow H'^0 H'^+$ and $pp 
\rightarrow H'^0  H'^-$, respectively.  The production cross section is 
parametrized as $\sigma (pp \rightarrow H'^0 H'^\pm) = 1.027 \times 
10^{-4} (\text{TeV}/m_{H'})^{4.2}$ pb,  where $m_{H'^0} = 
m_{H'^\pm} (\equiv m_{H'})$.
Since the branching ratio for $H'^0 \rightarrow b \bar b S'$ has a complicated 
dependency on model parameters, we take it as a free parameter. When charged 
Higgs boson is much heavier than $W^\pm$, $H'^\pm$ dominantly decays into 
$W^\pm$ and $S'$, and hence the branching ratio of $H'^\pm$ into jets and 
$S'$ is BR($H'^\pm \rightarrow \text{jets}+S'$) $\simeq$ 67.60 \%. Thus 
5$\sigma$ discovery reach for them is estimated as follows, 
\begin{equation}
 \begin{split}
   N_{\text{signal}} &= 
   \sigma (pp \rightarrow H'^0 H'^\pm) \times \mathcal{L}  \\[1mm]
   &~~\times \text{BR} (H'^0 \rightarrow b \bar b S') 
   \times \text{BR} (H'^\pm \rightarrow \text{jet} + S') 
   \times (b \text{-tag efficiency})   \\
   &\simeq  0.174 \left( \frac{\text{TeV}}{m_{H'}} \right)^{4.2} 
   \left( \frac{\mathcal{L}}{100 \text{ fb}^{-1}} \right) 
   \left( \frac{\text{BR}(H'^0 \rightarrow b \bar b S')}{10\%} \right) 
   \left( \frac{b\text{-tag efficiency}} {50\% \times 50\%} \right)   \\
   &> 5 \sqrt{N_{\text{BG}}} ~.
 \end{split}     \label{5sigma_2}
\end{equation}
The cross section of the $t \bar t$ background for these events is given by 
\cite{Baer:2000pe}, $\sigma_{t \bar t} = 0.89$ fb ($\sigma_{t \bar t} 
= 0.72$ fb) for the 2\% (1\%) $b$ mistagging probability.  
Thus assuming $\mathcal{L} = 100 \text{fb}^{-1}$,  BR($H'^0 \rightarrow 
b \bar b S'$) = 10\%, $b$-tag efficiency = $50\%\times 50 \%$, and demanding 
 the $t \bar t$ background 1 fb conservatively, the extra doublets
for $m_{H'^0} = m_{H'^\pm} \lesssim 260$ GeV would be discovered at the 
5$\sigma$ level.


Finally, we discuss the collider signature of the color-triplet states, introduced in 
order to maintain the successful gauge coupling unification, which is another key 
ingredient for the selection at the LHC.

Those colored extra particles are produced, and then they hadronize in the 
detector materials. 
The hadronized particles would be electrically either neutral or charged, and 
they can reverse the sign of its charge through the scattering in the detector 
materials \cite{Fairbairn:2006gg, Cheung:2002uz, Cheung:2003um}. 
If a hadronized particle is electrically neutral and would not undergo the charge 
reversal, it leaves detectors. The resultant large missing energy without energetic 
tau lepton or photon would be the mark for the {selection from ordinary} GMSB 
model. 
On the other hand, some of the hadronized particles are charged, and they lose 
their kinetic energy through ionization with the detector materials. The ionization 
energy loss is a function of $\beta \gamma$ and the electric charge of penetrating 
particle \cite{Amsler:2008zzb}.    When the energy loss and momentum of 
penetrating particle is measured, $\beta \gamma$ can be obtained, and hence its 
mass is determined. 
In addition, since massive long-lived charged particles produce a track in 
detectors, we can speculate its production rate. The estimated discovery 
potential with this method, however, is necessarily dependent on the scattering 
model, which of charge reversal predictions \cite{Fairbairn:2006gg}. Thus the 
discussion of feasibility for the discovery requires model dependent full analysis, and 
we leave it for future work.

\section{Conclusions} \label{sec:conc}

In this paper we proposed an extension of the GMSB model by adding extra doublets 
and singlet with parity odd under an additional $Z_2$ parity, while ordinary MSSM 
fields are parity-even. In this class of model, a natural candidate for DM appears as 
the lightest linear combination of additional $Z_2$-odd fields.
It has sizable interactions with nucleons and will be detected future/on-going direct 
detection experiments. We have also discussed typical collider signatures of this model 
and found that they may be discovered at 5 sigma level with an integrated luminosity 
of 100 ${\rm fb^{-1}}$, depending on model parameters.

The calculation of the relic abundance of bosonic-singlet DM indicates DM mass 
is around 250 GeV. If this is realized, this model could be confirmed by both the LHC 
experiment and future DM direct detection experiments. Furthermore, these DM can 
reproduce the DM-like events at CDMS-II. On the other hand, fermionic-doublet DM 
mass is around 1 TeV as shown in Fig. 1. In this case, it is very hard to discover these 
extra particles at the LHC, and DM direct detection experiments would not observe its 
signal.

\section*{Acknowledgment}

K.N. would like to thank the Japan Society for the Promotion of
Science for financial support.  The work was supported in part by the
Grant-in-Aid for the Ministry of Education, Culture, Sports, Science,
and Technology, Government of Japan, No. 20244037 and No. 2054252
(J.H.), No. 21111006 (K.N.) and No. 20007555 (M.Y.). 
The work of J.H. is also supported by the World Premier International Research Center Initiative (WPI Initiative), MEXT, Japan.


{}


\begin{thebibliography}{99}

\bibitem{Dine:1993yw}
  M.~Dine and A.~E.~Nelson,
  Phys.\ Rev.\  D {\bf 48}, 1277 (1993)
  [arXiv:hep-ph/9303230];
  M.~Dine, A.~E.~Nelson and Y.~Shirman,
  Phys.\ Rev.\  D {\bf 51}, 1362 (1995)
  [arXiv:hep-ph/9408384]; 
  M.~Dine, A.~E.~Nelson, Y.~Nir and Y.~Shirman,
  Phys.\ Rev.\  D {\bf 53}, 2658 (1996)
  [arXiv:hep-ph/9507378].
  
  
  
\bibitem{Giudice:1998bp}
  For a review, see G.~F.~Giudice and R.~Rattazzi,
  Phys.\ Rept.\  {\bf 322}, 419 (1999)
  [arXiv:hep-ph/9801271].
  
\bibitem{Moroi:1993mb}
  T.~Moroi, H.~Murayama and M.~Yamaguchi,
  Phys.\ Lett.\  B {\bf 303}, 289 (1993).
  
\bibitem{Fukugita:1986hr}
  M.~Fukugita and T.~Yanagida,
  Phys.\ Lett.\  B {\bf 174}, 45 (1986).
  
\bibitem{Viel:2005qj}
  M.~Viel, J.~Lesgourgues, M.~G.~Haehnelt, S.~Matarrese and A.~Riotto,
  Phys.\ Rev.\  D {\bf 71}, 063534 (2005)
  [arXiv:astro-ph/0501562].

\bibitem{Ichikawa:2009ir}
  K.~Ichikawa, M.~Kawasaki, K.~Nakayama, T.~Sekiguchi and T.~Takahashi,
  JCAP {\bf 0908}, 013 (2009)
  [arXiv:0905.2237 [astro-ph.CO]].
  
\bibitem{Kim:1986ax}
  For a review, see J.~E.~Kim,
  Phys.\ Rept.\  {\bf 150}, 1 (1987).
  
\bibitem{Rajagopal:1990yx}
  K.~Rajagopal, M.~S.~Turner and F.~Wilczek,
  Nucl.\ Phys.\  B {\bf 358}, 447 (1991).
  
\bibitem{Chun:1995hc}
  E.~J.~Chun and A.~Lukas,
  Phys.\ Lett.\  B {\bf 357}, 43 (1995)
  [arXiv:hep-ph/9503233].
    
\bibitem{Covi:2001nw}
  L.~Covi, H.~B.~Kim, J.~E.~Kim and L.~Roszkowski,
  JHEP {\bf 0105}, 033 (2001)
  [arXiv:hep-ph/0101009];
  A.~Brandenburg and F.~D.~Steffen,
  JCAP {\bf 0408}, 008 (2004)
  [arXiv:hep-ph/0405158].
  
\bibitem{Asaka:1998xa}
  T.~Asaka and M.~Yamaguchi,
  Phys.\ Rev.\  D {\bf 59}, 125003 (1999)
  [arXiv:hep-ph/9811451].



\bibitem{Kawasaki:2007mk}
  M.~Kawasaki, K.~Nakayama and M.~Senami,
  JCAP {\bf 0803}, 009 (2008)
  [arXiv:0711.3083 [hep-ph]].







\bibitem{Dimopoulos:1996gy}
  S.~Dimopoulos, G.~F.~Giudice and A.~Pomarol,
  Phys.\ Lett.\  B {\bf 389}, 37 (1996)
  [arXiv:hep-ph/9607225].

\bibitem{Hamaguchi:2007rb}
  K.~Hamaguchi, S.~Shirai and T.~T.~Yanagida,
  Phys.\ Lett.\  B {\bf 654}, 110 (2007)
  [arXiv:0707.2463 [hep-ph]].



\bibitem{Nelson:1983zb}
  A.~E.~Nelson,
  Phys.\ Lett.\  B {\bf 136}, 387 (1984).

\bibitem{Barr:1984fh}
  S.~M.~Barr,
  Phys.\ Rev.\  D {\bf 30},  1805 (1984).

\bibitem{Mahbubani:2005pt}
  R.~Mahbubani and L.~Senatore,
  Phys.\ Rev.\  D {\bf 73} (2006) 043510
  [arXiv:hep-ph/0510064].



\bibitem{Yanagida:1997yf}
  T.~Yanagida,
  Phys.\ Lett.\  B {\bf 400}, 109 (1997)
  [arXiv:hep-ph/9701394].

\bibitem{mgm}
  M.~Dine, Y.~Nir and Y.~Shirman,
  Phys.\ Rev.\  D {\bf 55},  1501 (1997)
  [arXiv:hep-ph/9607397]; 
  R.~Rattazzi and U.~Sarid,
  Nucl.\ Phys.\  B {\bf 501},  297  (1997)
  [arXiv:hep-ph/9612464].

\bibitem{Hisano:2007ah}
  J.~Hisano and Y.~Shimizu,
  Phys.\ Lett.\  B {\bf 655}, 269 (2007)
  [arXiv:0706.3145 [hep-ph]].
  
\bibitem{Komatsu:2010fb}
  E.~Komatsu {\it et al.},
  arXiv:1001.4538 [astro-ph.CO].
  
\bibitem{Griest:1991ma}
 K.~Griest and D.~Seckel,
Phys.\ Rev.\  D {\bf 43},  3191  (1991).


  
   
\bibitem{Arkani-Hamed:2006fe}
N.~Arkani-Hamed, A.~Delgado and G.F.~Giudice,
Nucl.\ Phys.\  B {\bf 741},  108  (2006)  
   
  
  
\bibitem{Kang:2006yd}
  J.~Kang, M.~A.~Luty and S.~Nasri,
  JHEP {\bf 0809}, 086 (2008)
  [arXiv:hep-ph/0611322].

\bibitem{Kusakabe:2009jt}
  M.~Kusakabe, T.~Kajino, T.~Yoshida and G.~J.~Mathews,
  Phys.\ Rev.\  D {\bf 80} (2009) 103501
  [arXiv:0906.3516 [hep-ph]].
  
\bibitem{Drees:1993bu}
  M.~Drees and M.~Nojiri,
  Phys.\ Rev.\  D {\bf 48}, 3483 (1993)
  [arXiv:hep-ph/9307208].

\bibitem{Jungman:1995df}
  G.~Jungman, M.~Kamionkowski and K.~Griest,
  Phys.\ Rept.\  {\bf 267}, 195 (1996)
  [arXiv:hep-ph/9506380].
  
\bibitem{Engel:1989ix}
  J.~Engel and P.~Vogel,
  Phys.\ Rev.\  D {\bf 40}, 3132 (1989).
  
\bibitem{Shifman:1978zn}
  M.~A.~Shifman, A.~I.~Vainshtein and V.~I.~Zakharov,
  Phys.\ Lett.\  B {\bf 78}, 443 (1978).
  
\bibitem{Gasser:1990ce}
  J.~Gasser, H.~Leutwyler and M.~E.~Sainio,
  Phys.\ Lett.\  B {\bf 253}, 252 (1991).

\bibitem{Adams:1995ufa}
  D.~Adams {\it et al.}  [Spin Muon Collaboration],
  Phys.\ Lett.\  B {\bf 357}, 248 (1995).
  
\bibitem{Ohki:2008ff}
  H.~Ohki {\it et al.},
  Phys.\ Rev.\  D {\bf 78}, 054502 (2008)
  [arXiv:0806.4744 [hep-lat]].
 
  
\bibitem{Ahmed:2008eu}
  Z.~Ahmed {\it et al.}  [CDMS Collaboration],
  Phys.\ Rev.\ Lett.\  {\bf 102}, 011301 (2009)
  [arXiv:0802.3530 [astro-ph]].

\bibitem{Ahmed:2009zw}
  Z.~Ahmed {\it et al.}  [The CDMS-II Collaboration],
  arXiv:0912.3592 [astro-ph.CO].
 
\bibitem{Angle:2008we}
   J.~Angle {\it et al.},
   Phys.\ Rev.\ Lett.\  {\bf 101}, 091301 (2008)
   [arXiv:0805.2939 [astro-ph]].
    
\bibitem{Griest:1986yu}
  K.~Griest and D.~Seckel,
  Nucl.\ Phys.\  B {\bf 283}, 681 (1987)
  [Erratum-ibid.\  B {\bf 296}, 1034 (1988)].
  
\bibitem{Ritz:1987mh}
  S.~Ritz and D.~Seckel,
  Nucl.\ Phys.\  B {\bf 304}, 877 (1988).
      
\bibitem{Kamionkowski:1991nj}
  M.~Kamionkowski,
  Phys.\ Rev.\  D {\bf 44}, 3021 (1991).
  
\bibitem{Jungman:1994jr}
  G.~Jungman and M.~Kamionkowski,
  Phys.\ Rev.\  D {\bf 51}, 328 (1995)
  [arXiv:hep-ph/9407351];
  M.~Kamionkowski, K.~Griest, G.~Jungman and B.~Sadoulet,
  Phys.\ Rev.\ Lett.\  {\bf 74}, 5174 (1995)
  [arXiv:hep-ph/9412213].

\bibitem{Desai:2004pq}
  S.~Desai {\it et al.}  [Super-Kamiokande Collaboration],
  Phys.\ Rev.\  D {\bf 70}, 083523 (2004)
  [Erratum-ibid.\  D {\bf 70}, 109901 (2004)]
  [arXiv:hep-ex/0404025].
  
\bibitem{Ackermann:2005fr}
  M.~Ackermann {\it et al.}  [AMANDA Collaboration],
  Astropart.\ Phys.\  {\bf 24}, 459 (2006)
  [arXiv:astro-ph/0508518].
  
\bibitem{Abbasi:2009uz}
  R.~Abbasi {\it et al.}  [IceCube Collaboration],
  Phys.\ Rev.\ Lett.\  {\bf 102}, 201302 (2009)
  [arXiv:0902.2460 [astro-ph.CO]].


\bibitem{Baer:1994nr}
  H.~Baer, C.~h.~Chen, F.~Paige and X.~Tata,
  Phys.\ Rev.\  D {\bf 50},  4508 (1994)
  [arXiv:hep-ph/9404212].
  
\bibitem{Baer:1995va}
  H.~Baer, C.~h.~Chen, F.~Paige and X.~Tata,
  Phys.\ Rev.\  D {\bf 53},  6241 (1996)
  [arXiv:hep-ph/9512383].
  
\bibitem{Aad:2009wy}
  G.~Aad {\it et al.}  [The ATLAS Collaboration],
  arXiv:0901.0512 [hep-ex].

\bibitem{Pukhov:2004ca}
  A.~Pukhov,
  arXiv:hep-ph/0412191.

\bibitem{Pumplin:2002vw}
  J.~Pumplin, D.~R.~Stump, J.~Huston, H.~L.~Lai, P.~Nadolsky and W.~K.~Tung,
  JHEP {\bf 0207},  012 (2002)
  [arXiv:hep-ph/0201195].

\bibitem{Baer:2000pe}
  H.~Baer, P.~G.~Mercadante, X.~Tata and Y.~l.~Wang,
  Phys.\ Rev.\  D {\bf 62},  095007  (2000)
  [arXiv:hep-ph/0004001].

\bibitem{Fairbairn:2006gg}
  M.~Fairbairn, A.~C.~Kraan, D.~A.~Milstead, T.~Sjostrand, P.~Z.~Skands and T.~Sloan,
  Phys.\ Rept.\  {\bf 438}, 1 (2007)
  [arXiv:hep-ph/0611040], and references therein.

\bibitem{Cheung:2002uz}
  K.~Cheung and G.~C.~Cho,
  Phys.\ Rev.\  D {\bf 67}, 075003 (2003)
  [arXiv:hep-ph/0212063].
  
\bibitem{Cheung:2003um}
  K.~Cheung and G.~C.~Cho,
  Phys.\ Rev.\  D {\bf 69} 017702 (2004)
  [arXiv:hep-ph/0306068].  

\bibitem{Amsler:2008zzb}
  C.~Amsler {\it et al.}  [Particle Data Group],
  Phys.\ Lett.\  B {\bf 667},1 (2008).
  

\end{thebibliography}
\end{document}